\documentclass[aps,twocolumn,floats,pre]{revtex4}
\pdfoutput=1
\usepackage{graphics,graphicx,epsfig}
\usepackage{amssymb}
\usepackage{epsf,epstopdf,wrapfig}

\newcommand {\e}[1]{\mathrm{~#1}}    
\newcommand {\vek}[1]{\mathbf{#1}}

\begin{document}

\title{Diffusion, dimensionality and noise in transcriptional regulation}

\author{Ga\v{s}per Tka\v{c}ik and William Bialek$^*$}

\affiliation{Joseph Henry Laboratories of Physics,  Lewis--Sigler Institute for Integrative Genomics, and $^*$Princeton Center for Theoretical Physics, Princeton University, Princeton, New Jersey 08544\\
Center for Studies in Physics and Biology, Rockefeller University, 1230 York Avenue, New York, New York 10065}

\begin{abstract}
The precision of biochemical signaling is limited by randomness in the diffusive arrival of molecules at their targets.  For proteins binding to the specific sites on the DNA and regulating transcription, the ability of the proteins to diffuse in one dimension by sliding along the length of the DNA, in addition to their diffusion in bulk solution, would seem to generate a larger target for DNA binding, consequently reducing the noise in the occupancy of the regulatory site.  Here we show that this effect is largely cancelled by the enhanced temporal correlations in one dimensional diffusion.  With realistic parameters, sliding along DNA has surprisingly little effect on the physical limits to the precision of transcriptional regulation.
\end{abstract}

\date{\today}

\maketitle
\section{Introduction}

Cells constantly regulate the expression levels of their genes.  A central motif in this regulatory process is the binding of transcription factor proteins to specific sites along the DNA.  The precision of transcriptional regulation is limited, ultimately, by randomness in the arrival of transcription factor (TF) molecules at these sites \cite{limit1}.  But proteins can find their binding sites on DNA by two very different mechanisms---either by diffusing in three dimensions through the surrounding solution, or by binding weakly and diffusing in one dimension along the contour of the DNA molecule.  The idea of diffusion or sliding along DNA goes back (at least) to the realization that the {\em lac} repressor seems to bind more rapidly to its target site than would be allowed by three dimensional diffusion alone \cite{riggs+al_70,berg+al_81}.  More recently,  the discussion of 3D vs. 1D diffusion has been revitalized by  theoretical analysis of the optimal search strategies  \cite{slutsky+mirny_04}, by new biochemical measurements \cite{gowers+al_05}, and by direct physical observations of the sliding motion \cite{Kabata+al_93,wang+al_06, elf+al_07}.  Here we consider the impact of dimensionality and diffusion on the physical limits to the precision of transcriptional regulation.

Any physical system which responds to the concentration of a signaling molecule will exhibit noise due to the random diffusion of these molecules in and out of its ``sensitive volume'' \cite{limit1}; the larger the volume, the smaller the (fractional) noise.  This leads to the intuition that 1D diffusion will have a huge impact on the noise because it effectively increases the size of the target to which the transcription factor is binding \cite{intuition}. Our main result  is that this intuition is wrong.  The problem is
that diffusion in one dimension has a very different statistical structure than in three dimensions, and so one cannot simply say that 1D sliding generates a larger 3D target.  We show that realistic combinations of 3D and 1D diffusion in fact have a surprisingly small effect on the limiting noise level---the increased size of the target is largely cancelled by stronger temporal correlations, which means that integrating over a fixed time gives fewer independent samples.  

\section{Concentration fluctuations}

Before starting on the problem of proteins binding to DNA, it is useful to recall some facts about concentration fluctuations for molecules free in solution \cite{LL}.  If we write the concentration as $c(\vek{x} , t)$, then the power spectrum of fluctuations $S_c(\vek{k},\omega )$ is defined by
\begin{eqnarray}
\langle \delta c(\vek{x} , t) \delta c(\vek{x'} , t')\rangle&& \nonumber\\
= \int {{d^d\vek{k}}\over {(2\pi )^d}}&&
{\hskip - 15pt}\int {{d\omega}\over{2\pi}}
S_c(\vek{k},\omega )
e^{ +i\vek{k}{\bf\cdot}(\vek{x} - \vek{x'}) -i\omega(t-t') },
\end{eqnarray}
where $d$ is the dimensionality of space.  For molecules  present at the mean concentration $\bar c$ and  diffusing freely with diffusion constant $D$, the power spectrum is given by
\begin{equation}
S_c(\vek{k},\omega ) = {{2 \bar c Dk^2}\over{(Dk^2)^2 + \omega^2}} ,
\end{equation}
where $k = |\vek{k}|$.  This result is independent of the dimensionality $d$, although of course the units of concentration are different in different dimensions.  If we integrate over all frequencies, which corresponds to making instantaneous measurements, we find a spatial power spectrum
\begin{equation}
\int {{d\omega}\over{2\pi}} S_c(\vek{k},\omega ) =  \bar c  .
\end{equation}
This is spatial white noise, and embodies the fact that a snapshot of molecules in solution will reveal a random, Poisson distribution: the variance of the molecule number in any finite volume is equal to the mean, so the power spectrum of concentration fluctuations is equal to the mean concentration, and again this is independent of dimensionality.

Consider a measurement of concentration that is averaged over some small region of space,
\begin{equation}
C(t) = \int d^d\vek{x} \, W(\vek{x}) c(\vek{x}, t) ,
\end{equation}
where $W$ is a weighting or windowing function that defines the size of the region we are interested in.  To interpret $C$ as an average of the local variable $c$, we must have $\int d^d \vek{x} W(\vek{x}) = 1$. The temporal fluctuations in $C$ are themselves determined by a power spectrum $S_C(\omega )$,
\begin{eqnarray}
\langle \delta C(t) \delta C(t') \rangle &=& \int {{d\omega}\over{2\pi}} S_C(\omega ) e^{-i\omega (t-t')}\\
S_C(\omega ) &=&\int {{d^d\vek{k}}\over{(2\pi )^d}} S_c(\vek{k},\omega ) {\large |} \tilde W (\vek{k}){\large |}^2 ,
\end{eqnarray}
where
\begin{equation}
\tilde W (\vek{k}) = \int d^d \vek{x} \,W(\vek{x})e^{i\vek{k}{\bf \cdot}\vek{x}} .
\end{equation}
The normalization of $W$ implies that $\tilde W (\vek{k} =0) = 1$.  If the size of the region is $\ell^d$, then $\tilde W (\vek{k})$ will fall to zero for $k$ much larger than $\Lambda \sim  1/\ell$.  Thus we can approximate
\begin{eqnarray}
S_C(\omega ) &=& \int {{d^d\vek{k}}\over{(2\pi )^d}} {{2 \bar c Dk^2}\over{(Dk^2)^2 + \omega^2}} {\large |} \tilde W (\vek{k}){\large |}^2\\
&\sim& \int_0^{\Lambda} dk\,k^{d-1} {{2 \bar c Dk^2}\over{(Dk^2)^2 + \omega^2}} .
\label{SCLambda}
\end{eqnarray}

We are interested in the behavior of the power spectrum at low frequencies, corresponding to measurements with long averaging times.  If we try simply to set $\omega =0$, then we find
\begin{equation}
S_C(\omega =0 ) \sim \int_0^{\Lambda} dk\,k^{d-1} {{2 \bar c }\over{Dk^2}} .
\end{equation}
For $d=3$, 
\begin{equation}
S_C^{3d}(\omega =0 ) \sim \int_0^{\Lambda} dk\,k^{2} {{2 \bar c }\over{Dk^2}} \sim {{2\bar c \Lambda}\over D} \sim {{\bar c}\over{D\ell}}.
\end{equation}
This corresponds to white noise in the time domain, so we expect that averaging over time will reduce the noise variance in proportion to the averaging time. More precisely, if we average for a time $\tau_{\rm int}$, we will be sensitive to frequencies $|\omega| < 1/\tau_{\rm int}$, so we will see a variance
\begin{eqnarray}
\langle (\delta C)^2\rangle_{3d} &\sim& \int_{|\omega| < 1/\tau_{\rm int}} {{d\omega}\over{2\pi}}S_C^{3d}(\omega )\\
&\sim&  \int_{|\omega| < 1/\tau_{\rm int}} {{d\omega}\over{2\pi}}S_C^{3d}(\omega =0)\\
&\sim& {{\bar c}\over{D\ell\tau_{\rm int}}}.
\label{BP_from_D_alone}
\end{eqnarray}
Although we have not done a full calculation including the effects of binding and unbinding to target sites, this is essentially the ``noise floor'' for a system which senses the concentration of signaling molecules by having them bind to a site of size $\ell$ \cite{limit1,bialek+setayeshgar_05,bialek+setayeshgar_06}.

In contrast, for $d=1$ we have 
\begin{equation}
S_C^{1d}(\omega =0 ) \sim \int_0^{\Lambda} dk\,{{2 \bar c }\over{Dk^2}},
\end{equation}
which is infrared divergent.  To get the right low frequency behavior in one dimension we have to be a bit more careful.  We have
\begin{equation}
S_C^{1d}(\omega ) \sim \int_0^{\Lambda} dk {{2 \bar c Dk^2}\over{(Dk^2)^2 + \omega^2}} .
\end{equation}
This integral is actually finite as $\Lambda\rightarrow\infty$, so we can write
\begin{eqnarray}
S_C^{1d}(\omega ) &\sim& \int_0^\infty dk {{2 \bar c Dk^2}\over{(Dk^2)^2 + \omega^2}} \\
&=& {{2 \bar c}\over{\sqrt{D\omega}}} \int_0^\infty d\left(k \sqrt{D\over \omega}\right){{Dk^2/\omega}\over{(Dk^2 /\omega )^2 +1}}\\
&=&  {{2 \bar c}\over{\sqrt{D\omega}}} \int_0^\infty dq{{q^2}\over{q^4 +1}}\\
&\sim& {{\bar c}\over\sqrt{D\omega}} .
\end{eqnarray}
Now the variance of measurements averaged over a time $\tau_{\rm int}$ becomes
\begin{eqnarray}
\langle (\delta C)^2\rangle_{1d} &\sim& \int_{|\omega| < 1/\tau_{\rm int}} {{d\omega}\over{2\pi}}S_C^{1d}(\omega )\\
&\sim&  \int_{|\omega| < 1/\tau_{\rm int}} {{d\omega}\over{2\pi}}{{\bar c}\over\sqrt{D\omega}}\\
&\sim& {{\bar c}\over\sqrt{D\tau_{\rm int}}}.
\label{BP1D_from_D_alone}
\end{eqnarray}
We see that in one dimension, the {\em variance} of concentration measurements declines only as the square root of the measurement time.  This is contrast to the three dimensional case, where the standard deviation of the noise declines as the square root of time but the variance declines in direct proportion to the measurement time, as seen in Eq (\ref{BP_from_D_alone}).

\section{Binding with 3D diffusion}
If transcription factors bind  specific sites on the DNA to control the rate of gene expression, then the noise in the regulated gene product will contain a contribution from the noise in the occupancy of the specific site. Here we want to compute  this noise contribution, starting with the case where the transcription factors find their cognate site by 3D diffusion alone and progressing towards more complicated translocation strategies in the following sections.

In this section we briefly review the calculations of Ref \cite{bialek+setayeshgar_05}. Consider a cell volume with an average concentration $\bar{c}$ of TFs present, and a binding site located at $\vek{x}_0=0$. The transcription factor molecules can diffuse with a bulk diffusion constant $D_3$, bind to the specific site at a rate $k_+ c$, and dissociate back into the free solution at a rate $k_-$, according to the following set of equations:
\begin{eqnarray}
\frac{\partial c}{\partial t}&=&D_3\nabla^2 c - \dot{n}\,\delta(\vek{x}-\vek{x}_0), \label{e3d1} \\
\dot{n}&=&k_+ c(\vek{x}_0,t)(1-n)-k_-n. \label{e3d2}
\end{eqnarray}
To determine the fluctuations in the binding site occupancy $n$, we linearize the equations about the mean occupancy $n(t)=\delta n(t)+\bar{n}$ with $\bar{n}=k_+\bar{c}/(k_+\bar{c}+k_-)$, and around the mean concentration, $c(\vek{x},t)=\delta c(\vek{x},t)+\bar{c}$, and write the perturbations as Fourier modes:
\begin{eqnarray}
\delta n(t)&=&\int\frac{d\omega}{(2\pi)}  e^{-i\omega t}\, \delta\tilde{n}(\omega),	\\
\delta c(\vek{x},t)&=&\int \frac{d\omega}{(2\pi)}\int\frac{d^3{k}}{(2\pi)^3}e^{-i(\omega t + \vek{k}{\bf \cdot}\vek{x})}\,\delta\tilde{c}_{\vek{k}}(\omega).
\end{eqnarray}
 We are looking for the power spectrum of fluctuations in occupancy, which we can define by
 \begin{equation}
 \langle \delta \tilde{n}(\omega)\delta \tilde{n}^*(\omega ')\rangle
 = 2\pi \delta(\omega - \omega ') S_n(\omega ).
 \end{equation}

If there exists a low frequency limit, $S_n(\omega\rightarrow 0)$, and we average the noise for a time $\tau_{\rm int}$ that is longer than all other relevant timescales in the problem, we can [by analogy with the arguments leading to Eq (\ref{BP_from_D_alone})] use  it to calculate the observable variance by computing $\sigma_n^2=S_n(\omega \rightarrow 0)/\tau_{\rm int}$.

Having introduced the necessary notation, we can compute the fluctuations in concentration \emph{at} the binding site, $\vek{x}_0=0$, from Eq (\ref{e3d1}):
\begin{equation}
\delta \tilde{c}(\vek{x}_0,\omega)=i\omega \delta \tilde{n} \int \frac{d^3{k}}{(2\pi)^3}\frac{1}{-i\omega+D_3 k^2}.
\label{e3d2b}
\end{equation}
This integral is ultraviolet divergent, and must be cut off at $\Lambda=\frac{\pi}{a}$, where $a$ is the binding site size; keeping leading terms in $i\omega$ one obtains:
\begin{equation}
\delta \tilde{c}(\vek{x}_0,\omega\rightarrow 0)=\frac{i\omega\delta \tilde{n}}{2\pi D_3 a}. \label{e3d3}
\end{equation}
The linearization of Eq (\ref{e3d2}) yields
\begin{equation}
-i\omega \delta \tilde{n}=-\frac{\delta \tilde{n}}{\tau_c} + k_+(1-\bar{n})\delta \tilde{c}(\vek{x}_0,\omega) + k_-\bar{n}\beta\delta \tilde{F},
\label{e3d4}
\end{equation}
where $\tau_c^{-1}=k_+\bar{c}+k_-$ is the time scale of occupancy fluctuations, $\beta=1/k_BT$, and $\delta \tilde{F}$ is the fluctuation in 
the free energy difference between the bound and empty states of the site; it is the `force' that is thermodynamically conjugate to the `displacement' $\delta n$.
After substituting the local concentration fluctuations, Eq (\ref{e3d3}), into Eq (\ref{e3d4}), one can finally use the fluctuation--dissipation theorem to compute the power spectrum of the noise in occupancy:
\begin{equation}
S_n(\omega)=\frac{2k_B T}{\omega} \mathcal{I}m\,\frac{\delta \tilde{n}}{\delta \tilde{F}}. \label{psn}
\end{equation}
The result of Ref \cite{bialek+setayeshgar_05} was that, in the case of 3D diffusion to the binding site, the low-frequency limit of the occupancy noise power spectrum is:
\begin{equation}
S_n(\omega\rightarrow 0)=\frac{2\bar{n}(1-\bar{n})^2}{k_-}+\frac{\bar{n}^2(1-\bar{n})^2}{\pi D_3 a \bar{c}}. \label{e3d-nps}
\end{equation}
The first term corresponds to the binomial switching fluctuations as the occupancy of the specific site changes between full and empty; this term depends on the microscopic details of the TF-DNA interaction (here embodied by the off-rate constant $k_-$). The second term is caused by the fluctuating diffusive flux to the binding site, and provides a lower bound on the noise, independent of details. Expressed in terms of the equivalent concentration fluctuations, $S_c(\omega)=S_n(\omega)\left| \frac{d\bar{n}}{dc} \right|^{-2}$, the noise spectrum for the second term is 
\begin{equation}
S_c(\omega\rightarrow 0)=\bar{c}/\pi D_3 a, \label{e3d5}
\end{equation}
with the associated fractional variance
\begin{equation}
\left(\frac{\sigma_c}{\bar{c}}\right)^2=\frac{1}{\pi}\times\frac{1}{ D_3  \bar{c}\tau_{\rm int}}\times\frac{1}{a},
\label{e3d6}
\end{equation}
where, to facilitate comparison with results that follow, we explicitly pull out the leading numerical factor and $1/$(effective binding site size). This is the noise lower bound, consistent with the result of Eq (\ref{BP_from_D_alone}) for $\ell=a$, that depends solely on the rate of 3D diffusion $D_3$, the binding site size $a$, and the amount of time averaging that the system performs, $\tau_{\rm int}$. 
\section{Binding with 1D diffusion}
Consider the case where the diffusion in bulk does not occur, and the TF is only free to slide along the length of the DNA. Let $\xi(x,t)$ denote the 1D concentration along the DNA contour, which we imagine to be stretched in the $\hat{\vek{x}}$ direction, and let $D_1$ be the corresponding diffusion constant. Equations (\ref{e3d1},\ref{e3d2}) remain unchanged but for the new notation, i.e. $c(\vek{x},t)\rightarrow \xi(x,t), D_3\rightarrow D_1, \nabla^2\rightarrow \partial_x^2$; we will use $q$ to denote the Fourier variable conjugate to $x$. We again need to calculate the concentration fluctuations at the binding site (cf. Eq (\ref{e3d2b})):
\begin{equation}
\delta\tilde{\xi}(x_0,\omega)=i\omega\delta\tilde{n}\int_{-\infty}^{\infty}\frac{dq}{(2\pi)}\frac{1}{-i\omega+D_1q^2}.
\label{e1d2a}
\end{equation}
This integral diverges as $q \rightarrow 0$ if we set $\omega=  0$, but in contrast to the discussion above there is no problem as $q\rightarrow\infty$. The full result is that:
\begin{equation}
\delta\tilde{\xi}(x_0,\omega)=\frac{i\omega \delta \tilde{n}}{2 D_1 \sqrt{-i\omega/D_1}}. \label{e1d2}
\end{equation}
Note that unlike Eq (\ref{e3d3}), this result does not contain the binding site dimensions. Using Eq (\ref{e1d2}) we compute the diffusive contribution to the noise power spectrum (cf. Eq (\ref{e3d5}), but no limit $\omega\rightarrow 0$):
\begin{equation}
S_\xi(\omega)=\frac{\bar{\xi}}{\sqrt{2\omega D_1}}. 
\label{spec1d}
\end{equation}
The noise variance is obtained by integrating the noise power spectrum, as in the arguments leading to Eq (\ref{BP1D_from_D_alone}),
\begin{equation}
\left(\frac{\sigma_\xi}{\bar{\xi}}\right)^2=\frac{2}{\pi}\frac{1}{\sqrt{2D_1\tau_{\rm int}}\bar{\xi}}.
\end{equation}
We see that the noise variance declines as $1/\sqrt{\tau_{\rm int}}$, consistent with the simpler calculation of concentration fluctuations in Eq (\ref{BP1D_from_D_alone}).  This again is in contrast to our intuition that variances should decline as $1/\tau_{\rm int}$, which is the result for binding coupled to 3D diffusion.
The difference between 1D and 3D  can be understood by realizing that in 1D diffusion, a particle leaving the binding site at the origin has a large probability (in fact, probability 1) of returning back to the origin. A receptor trying to estimate the local concentration of TFs will therefore be unable to collect samples that are truly independent, and the variance in the measurements will consequently decrease at a rate that is slower than  expected from averaging over independent measurements.

Suppose that the DNA with the sliding TFs is embedded into the cytoplasm, where the bulk concentration of TFs is $\bar{c}$; the TFs can jump onto the DNA at a rate (per unit length) $\kappa_+ \bar{c}$, where we expect $\kappa_+ \sim D_3$.  The TF will stay on the strand for a  \emph{residence time}, $\tau_{\rm r}=\kappa_-^{-1}$, where $\kappa_-$ is the rate for dissociating from the DNA. There will be an equilibrium between the 1D and 3D concentrations, $\bar{c}=\bar{\xi}/\tau_{\rm r} \kappa_+\approx \bar{\xi}/\tau_{\rm r} D_3$. With these identifications, the noise variance can be rewritten in a form similar to the result in Eq (\ref{e3d6}):
\begin{equation}
\left(\frac{\sigma_c}{\bar{c}}\right)^2=\frac{\sqrt{2}}{\pi} \times \frac{1}{D_3 \bar{c} \tau_{\rm int} }\times \frac{1}{\sqrt{D_1\tau_{\rm r}^2/\tau_{\rm int}}}.
\label{e1d4}
\end{equation}
Here, the effective binding site size is $a_{\rm eff}=\sqrt{D_1\tau_{\rm r}}\sqrt{\tau_{\rm r}/\tau_{\rm int}}$. Naively, one could think that the effective site size would be equal to the average length that the TF explores during 1D diffusion on the DNA, $b=\sqrt{D_1\tau_{\rm r}}$, and therefore $a_{\rm eff}\sim b$, but that would be wrong:  the effective site size depends on $\tau_{\rm int}$ to compensate for the highly correlated fluctuations in 1D diffusion. Because $\tau_{\rm int}\gg \tau_r$ \cite{tavg}, the effective site size will be much smaller than $b$. The apparent reduction in the noise expected in the naive picture -- because the size of the binding site, $a$, is replaced by presumably much larger length explored by 1D diffusion, $b$ -- must in reality be traded off against the longer required integration time.  As a result, it is not immediately clear whether 1D diffusion decreases the noise lower bound compared to 3D case. 
\section{Combined 1D and 3D diffusion}
In the discussion so far we have been missing a parameter that would interpolate  between two qualitatively different noise regimes: the 3D result of Eq (\ref{e3d6}) and the 1D result of Eq (\ref{e1d4}). If the rate for dissociation from the sliding state into the bulk is increased, the 1D case must approach the 3D result; the residence time will grow ever shorter and will, at some scale, break the strong correlations reflective of the 1D sliding mode with its high probability for returning to the origin.

\begin{figure}[b]
  \begin{center}    
\hfill
\centerline{\includegraphics*[height=3in]{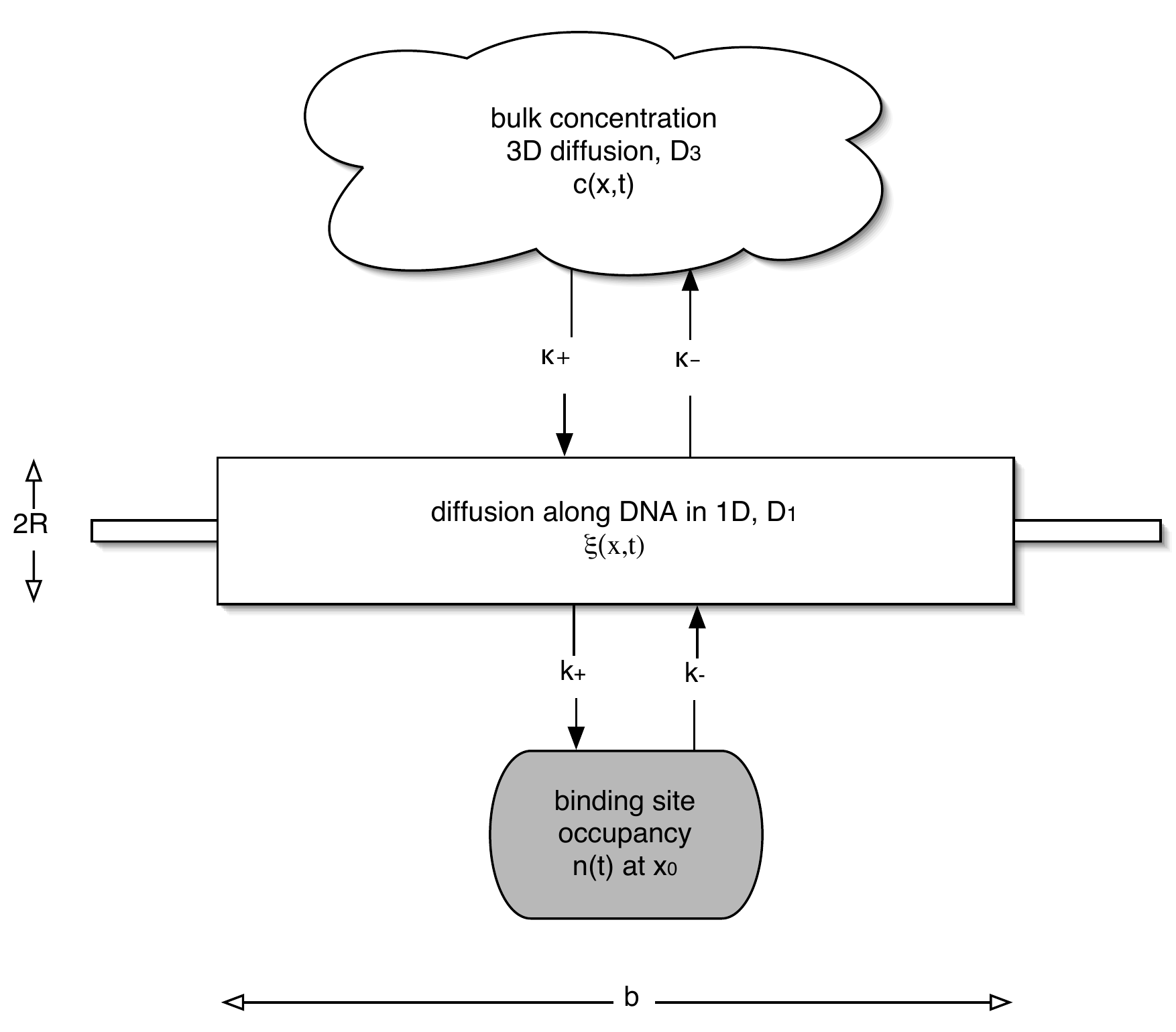}}
\caption[Schematic diagram of transcription factor translocation on the DNA.]{The transcription factors can either be free in solution at concentration $c$, or they can enter a region on the DNA where they diffuse by sliding. The 1D concentration is denoted by $\xi(x,t)$. The specific binding site on the DNA is at  location $x_0=0$;  $k_+\xi(x_0)$ and $k_-$ are the on-- and off--rates for transitions between the 1D sliding state and the bound state on the specific site; $\kappa_+c$ is the rate per unit length for transition onto the DNA from free solution, and $\kappa_-=\tau_{\rm r}^{-1}$ is the rate for dissociating from 1D solution into the 3D solution. The effective radius of the DNA molecule is $R$ and the ``sliding length,'' or the average distance along the contour covered in a single 1D random walk before dissociation, is denoted by $b$.}
\label{fig:sliding-noise-scheme}
  \end{center}
\end{figure}

\subsection{Perfect mixing in bulk solution}
We first solve a simplified case where the TF diffuses on the DNA, but if it dissociates into the bulk, it mixes very quickly and there is no correlation between the point of dissociation and subsequent re-association to the DNA strand. We expect that the divergence observed in the pure 1D case, Eq (\ref{e1d2a}), will now get regularized.
The dynamical equations for this case are:
\begin{eqnarray}
\frac{dn}{dt}&=&k_+\xi(x_0,t)(1-n)-k_-n	\\
\frac{\partial\xi}{\partial t}&=&D_1\frac{\partial^2\xi}{\partial x^2}-\dot{n}\delta(x-x_0)-\kappa_-\xi+\kappa_+ C	\\
\frac{dC}{dt}&=&\kappa_-\int dx\,\xi(x,t)-\kappa_+ L C
\end{eqnarray}
Here, $C(t)$ is the bulk concentration that is perfectly mixed and therefore does not have any spatial dependence; TFs in the bulk are enclosed into a box with a side of length $L$ and the DNA inside is an infinitely thin line of length $L$; and $\kappa_-$ and $\kappa_+$ are rates for transitioning between 1D concentration $\xi$ and the bulk $C$.  When linearized and written out in terms of their Fourier components, the equations couple to the zero mode, $\delta \tilde{\xi}(q=0,\omega)$, differently than to non-zero modes:
\begin{eqnarray}
(-i\omega+1/\tau_c)\delta\tilde{n}&=&k_+(1-\bar{n})\delta\tilde\xi(x_0)+k_-\bar{n}\beta\delta F \\
(-i\omega + D_1 q^2)\delta\tilde{\xi}_q&=&i\omega\delta\tilde{n} -\kappa_-\delta\tilde{\xi}_q+ \kappa_+L \delta \tilde{C}	\label{ee1}\\
(-i\omega + \kappa_+L)\delta\tilde{C}&=&\kappa_-\delta\tilde{\xi}_{q=0} \label{ee2}
\end{eqnarray}
We use Eq (\ref{ee1}) at $q=0$ together with Eq (\ref{ee2}) to express $\delta \tilde{C}$ and substitute it back into Eq (\ref{ee1}) for non-zero $q$; it is then straightforward to write $\delta\tilde{\xi}(x_0)$ as:
\begin{eqnarray}
\delta\xi(x_0)&=&\int\frac{dq}{(2\pi)}\frac{i\omega\Sigma(\omega)\,\delta \tilde{n}}{-i\omega + D_1q^2+\kappa_-}	\label{eei} \\
&=&\frac{i\omega\Sigma(\omega)}{2\sqrt{D_1}}\frac{\delta\tilde{n}}{\sqrt{\kappa_--i\omega}}, 
\end{eqnarray}
where 
 \begin{equation}
 \Sigma(\omega)=1+\frac{\kappa_-\kappa_+L}{(-i\omega+\kappa_-)(-i\omega+\kappa_+L)-\kappa_-\kappa_+L}.
 \end{equation}
Note that as $\omega\rightarrow 0$, $\Sigma(\omega)$ will diverge; before computing the power spectrum of fluctuations $S_n$ using Eq (\ref{psn}) we therefore need to expand Eq (\ref{eei}) to the leading power of $i\omega$. On the other hand, the momentum integral in Eq (\ref{eei}) now converges due to the $\kappa_-$ term in the denominator.

After some algebra, the effective power spectrum in the concentration fluctuations is:
\begin{equation}
S_\xi(\omega\rightarrow 0) = \frac{\bar{\xi}}{2\sqrt{D_1\kappa_-}}\left(1+\frac{1-\gamma}{(1+\gamma)^2}\right), \label{eecf}
\end{equation}
where $\gamma=\kappa_+L/\kappa_-$ is the ratio of on- and off-rates for going into 1D solution. The $\gamma$-dependent expression in parenthesis has a minimum of $7/8$ at $\gamma_0=3$, and therefore a noise floor exists that cannot be eliminated by some convenient combination of the rate parameters.
 
Rewritten in terms of the ``equivalent  concentration'' $\bar{c}=\bar{C}/L^3=\kappa_-\bar{\xi}/\kappa_+L^3\approx \kappa_-\bar{\xi}/D_3$, the concentration noise floor is (cf. Eq (\ref{e1d4})): 
\begin{equation}
\left(\frac{\sigma_c}{\bar{c}}\right)^2= \frac{7}{16}\times\frac{1}{D_3 \bar{c} \tau_{\rm int}}\times\frac{1}{\sqrt{D_1\tau_{\rm r}}}.
\end{equation}
This is the result we might naively have expected, namely that sliding along the DNA creates a larger target, of size $a_{\rm eff}\approx b$.
But we get this result 
\emph{only} because we have assumed that the TF can repeatedly dissociate into a well-mixed bulk solution, instantly losing the memory of the point along the DNA from which it dissociated.

\subsection{Diffusion in bulk solution}

We will now solve the coupled 3D-1D diffusion problem without assuming that in the bulk solution the transcription factors mix perfectly. The relevant quantities are schematized in Fig \ref{fig:sliding-noise-scheme}. 

\begin{widetext}

We describe the system by the following set of equations:
\begin{eqnarray}
\frac{dn}{dt}&=&k_+\xi(x_0,t)(1-n)-k_-n,	 \label{eq:ss-1}\\
\frac{\partial \xi(x,t)}{\partial t}&=&D_1\frac{\partial^2\xi(x,t)}{\partial^2 x}-\frac{dn}{dt}\delta(x-x_0)+
\kappa_+ c(x,\vek{R}=0,t)-\kappa_-\xi(x,t),   \label{eq:ss-2}\\
\frac{\partial c(x,\vek{R},t)}{\partial t}&=&D_3\nabla^2 c(x,\vek{R},t)-\left[\kappa_+c(x,\vek{R}=0,t)-\kappa_-\xi(x,t)\right]\delta(y)\delta(z).	 \label{eq:ss-3}
\end{eqnarray}
We have again assumed that DNA is stretched along the $x$-axis and that it is an infinitely thin molecule. $\xi$ is a function of only one variable, $x$, while $c$ is a function of coordinate $x$ and radial coordinates $\vek{R}$ that are perpendicular to the $\hat{\vek{x}}$ direction.
We linearize and Fourier transform the equations as follows:
\begin{eqnarray}
-i\omega \delta\tilde{n}&=&-\frac{\delta\tilde{n}}{\tau_c}+ k_+(1-\bar{n})\delta \tilde{\xi}(x_0)+ k_-\bar{n}\beta\delta  \tilde{F}\\
-i\omega \delta \tilde{\xi}_q&=&-D_1q^2\delta \tilde{\xi}_q+i\omega \delta \tilde{n}-\kappa_-\delta \tilde{\xi}_q+
\kappa_+ \delta \hat{c}_{q} \label{d31e2}\\
-i\omega \delta \tilde{c}_{q,\vek{k_\perp} }&=&-D_3\left(q^2+k_\perp^2\right) \delta \tilde{c}_{q,\vek{k_\perp}}+\kappa_-\delta \tilde{\xi}_q-\kappa_+\delta\hat{c}_q, \label{d31e21}
\end{eqnarray}
\end{widetext}
where again $q$ is the spatial Fourier variable conjugate to $x$ and $\vek{k_\perp}$ is conjugate to $\vek{R}$.
Note that $\delta\tilde{\xi}_q$ is function only of $q$, while
$\delta\tilde{c}_{q,\vek{k_\perp}}$ is a 3D Fourier transform of the bulk concentration fluctuations that depends on both $q$ and $\vek{k_\perp}$. Finally, $\delta\hat{c}(q,\vek{R}=0,\omega)$ are Fourier modes of concentration fluctuations along the $\hat{\vek{x}}$ direction, evaluated at the location of the DNA strand, $\vek{R}=0$. 

We first compute $\delta\hat{c}_q$ from Eq (\ref{d31e21}):
\begin{eqnarray}
\delta\hat{c}_q&=&\int\frac{d^2 k_\perp}{(2\pi)^2}\delta\hat{c}_{q,\vek{k_\perp}} \\
\delta\hat{c}_q&=&\int\frac{d^2 k_\perp}{(2\pi)^2}\frac{\kappa_-\delta\hat{\xi}_q-\kappa_+\delta\hat{c}_q}{-i\omega+D_3q^2+D_3 k_\perp^2}\\
\delta\hat{c}_q&=&\frac{\kappa_-\delta\hat{\xi}_q-\kappa_+\delta\hat{c}_q}{4\pi D_3}\log\left\{1+\frac{\Lambda^2}{k_0^2}\right\} \label{d31e4}
\end{eqnarray}

Here, $k_0^2=q^2-i\omega/D_3$, and $\Lambda=\frac{\pi}{R}$ is the ultraviolet cutoff at the radial size  of the DNA molecule. We can substitute $\delta\hat{c}_q$ from Eq (\ref{d31e4}) into Eq (\ref{d31e2}) to obtain the expression for $\delta\hat{\xi}(x_0)$:
\begin{eqnarray}
\delta \tilde{\xi}(x_0)&=&i\omega\delta \tilde{n}\int\frac{dq}{(2\pi)}\frac{1}{-i\omega+D_1q^2+\kappa_-F^{-1}(q)},
\label{eq:ss-int}\\
F(q)&=&1+\frac{\kappa_+}{4\pi D_3}\log\left\{1+\frac{\Lambda^2}{-i\omega/D_3+q^2}\right\}.
\end{eqnarray}
Compared to the pure 1D result, Eq (\ref{e1d2a}),  Eq (\ref{eq:ss-int}) now contains a new term $\kappa_-F^{-1}(q)$ in the denominator; for $\kappa_-=0$ this term vanishes and the result reverts to the 1D case as expected. $F(q)$ depends on the momentum $q$, whereas in the perfect mixing case, Eq (\ref{eei}), it was simply equal to 1. The integrand of Eq (\ref{eq:ss-int}) is still divergent for $\omega=0$ as $q\rightarrow 0$, but the integral nevertheless converges. Assuming a non-zero $\kappa_-$, the limit $\omega\rightarrow 0$ therefore exists, and we can integrate:
\begin{equation}
\delta\tilde{\xi}(x_0,\omega\rightarrow 0)=\frac{i\omega\delta \tilde{n}}{\pi\Lambda D_1} I(\alpha,\beta),
\end{equation}
where
\begin{eqnarray}
I(\alpha,\beta)&=&\int_0^\infty\frac{dt}{t^2+\beta\left(1+\alpha \log \left(1+t^{-2}\right)\right)^{-1}}, \label{eq:ss-inti}\\
\alpha&=&\frac{\kappa_+}{4\pi D_3},	\\
\beta&=&\frac{\kappa_-}{\Lambda^2 D_1}=\left(\frac{R}{\pi b}\right)^2.
\end{eqnarray}
The noise power spectrum for occupancy follows in close analogy to Eq (\ref{e3d-nps}):
\begin{equation}
S_n(\omega\rightarrow 0)=\frac{2 \bar{n}(1-\bar{n})^2}{k_-}+\frac{2\bar{n}^2(1-\bar{n})^2}{\pi \Lambda D_1\bar{\xi}} I(\alpha,\beta).
\end{equation}
The effective spectrum of local concentration fluctuations corresponding to the second term is (cf. Eqs (\ref{e3d5}, \ref{spec1d}, \ref{eecf})):
\begin{equation}
S_c(\omega\rightarrow 0)=\frac{2\bar{\xi} R}{\pi^2 D_1}I(\alpha,\beta).
\end{equation}
The associated noise variance is:
\begin{equation}
\left(\frac{\sigma_\xi}{\bar{\xi}}\right)^2=\frac{2I(\alpha,\beta)}{\pi^2 D_1 \tau_{\rm int}\bar{\xi}/R}.
\label{e31dd}
\end{equation}
With the above definition of $\alpha$, the equilibrium between 1D and 3D concentrations turns out to be $\bar{c}=\bar{\xi}/4\pi D_3\alpha \tau_{\rm r} $. The noise variance in concentration can be rewritten as:
\begin{equation}
\left(\frac{\sigma_c}{\bar{c}}\right)^2=\frac{\beta I(\alpha,\beta)}{2\pi \alpha}\times\frac{1}{D_3 \bar{c}\tau_{\rm int}}\times \frac{1}{R} \label{combres}
\end{equation}

\begin{figure}[b]
  \begin{center}    
\centerline{\includegraphics[height=2.2in]{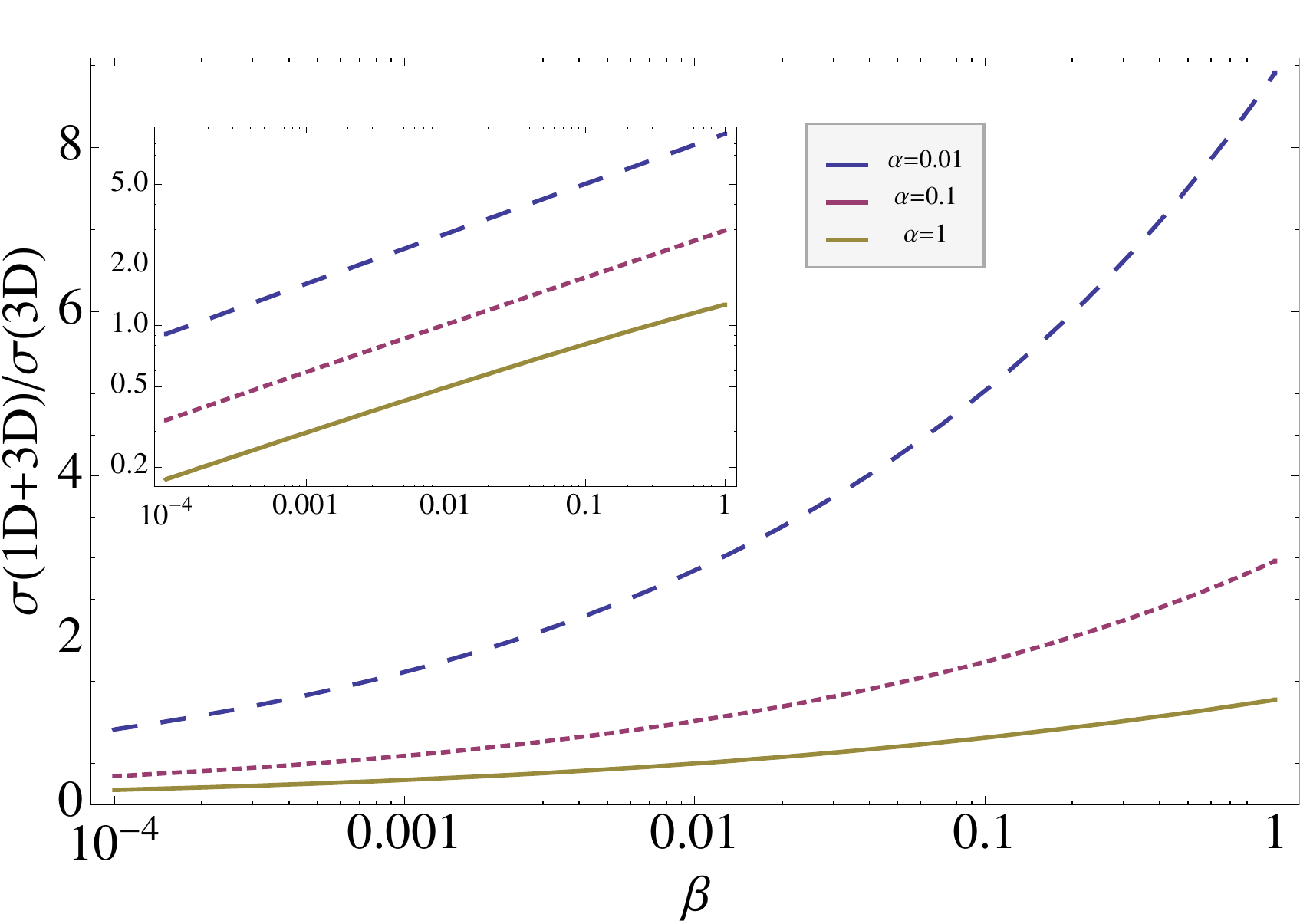}}
\caption[The effect of 1D diffusion along the DNA on the noise.]{The relative decrease in noise magnitude, compared to the pure 3D diffusion model, as a function of parameters $\alpha$ and $\beta$. Three values for $\alpha$ are shown, spanning two orders of magnitude ($\alpha=1$ solid, $\alpha=0.1$ dotted, $\alpha=0.01$ dashed); $\beta$ covers the relevant range if typical 1D diffusion length is as expected from search time optimality arguments (order hundred base pairs). }
\label{ffbeta}  
\end{center}  
\end{figure}

The result looks similar to the pure 3D diffusion case in Eq (\ref{e3d6}): the noise due to concentration fluctuations has its  length scale $a$ (receptor size in pure 3D case) replaced with $R$ (effective DNA cross-section in the combined 1D/3D case), and the noise contribution term gets multiplied by the prefactor that we examine next.

We expect the parameter $\alpha={\kappa_+}/{4\pi D_3}$ to be close to 1 for diffusion limited approach to DNA.  Imagine that the area that TF attempts to hit in order to bind non-specifically to the DNA is a cylindrical segment of DNA with radius $R$ and length $b$. If we were treating the cylindrical DNA segment as a sphere of radius $b$, then Smoluchowski on-rate limit $\sim 4\pi D_3 b$ would apply \cite{boundlinear}. In the first approximation, the on rate per unit length, $\kappa_+$, would then be $4\pi D_3$ and $\alpha=1$. Exact derivations for Smoluchowski limit in simple geometries were reviewed in Ref \cite{Berg+vonHippel_85}; for elongated objects (prolate ellipsoid with long semiaxis $b$ and short semiaxis $R$), the diffusion limited rate is given by $4\pi D_3 b/\ln(2b/R)$, and therefore $\alpha\sim \ln^{-1}(2b/R)$. $\alpha$ is hence less then one, and has a weak logarithmic dependence on the ratio of the scale of the persistence length and the effective radius of the DNA.

The parameter $\beta$ is approximately the square of the ratio between the cross-section of the 1D cylinder (the ``target'' that 3D diffusion has to hit) and the average ``sliding length'' $b$ along the DNA. $R$  must be of order of several nanometers; while $b$ is, at DNA stacking length of $a=0.3\e{nm}$ per base-pair and $100\e{bp}$ average diffusion length \citep{halford+marko_04, slutsky+mirny_04}, around $b=10-100\e{nm}$.  It is therefore not unreasonable to assume that the factor $\beta$ could be as low as $\beta\sim 10^{-3}-10^{-2}$. The corresponding decrease in noise relative to pure 3D diffusion  is shown in Fig \ref{ffbeta}.

Note that as $\beta\rightarrow \infty$ one should recover the pure 3D diffusion result. Looking back at expression for $I(\alpha,\beta)$ in Eq (\ref{eq:ss-inti}), as $\beta$ increases, the $t^2$ term in the denominator of the integrand is becoming irrelevant. If we neglect it completely, the integral is solvable analytically: 
\begin{equation}
I(\alpha,\beta\rightarrow \infty)=\beta^{-1}\left(\frac{\Lambda_x}{\Lambda} + \pi \alpha\right),
\end{equation}
where $\Lambda_x=\frac{\pi}{a}$ is the cut-off at the binding site size along the DNA, and $\Lambda_x/\Lambda=R/a$. Inserting the integral in the large $\beta$ limit into the noise result, Eq (\ref{combres}), we see that $\beta$ cancels and we get:
\begin{eqnarray}
\left(\frac{\sigma_c}{\bar{c}}\right)^2&=&\frac{1}{2\pi}\left(\pi+\frac{R}{\alpha\,a}\right)\times \frac{1}{D_3\bar{c}\tau_{\rm int}}\times \frac{1}{R}\\
&>& {1\over {2\pi \alpha}}\times 
\frac{1}{D_3\bar{c}\tau_{\rm int}}\times {1\over a} ,
\end{eqnarray}
which is essentially the 3D result, Eq (\ref{e3d6}).

\section{Discussion}
While the role of 1D diffusion of transcription factors along the DNA has been recognized as a possible explanation for the observed fast DNA-TF association rates, the impact of this additional mode of TF translocation on the noise in binding site occupancy remained unexplored. The question is important  for  two reasons: first, the diffusive contribution to the noise in gene expression must fundamentally limit the precision of transcriptional regulation; and second, there is an appealing argument that 1D diffusion could drastically increase the target area on the DNA that TFs have to find and correspondingly lower the limiting diffusive noise. Here we show that this intuitive argument is wrong---while there might be some reduction in the noise if the bulk diffusion is supplemented by 1D sliding, this reduction is not  expected to be significant.

Much has been said about possible TF translocation strategies on the DNA such as three-dimensional volume exchange, local dissociation-reassociation reactions (hopping), sliding along the DNA and intersegmental transfers mediated by DNA-looping \cite{berg+al_81,Shimamoto_99}, and about the ways in which these mechanisms influence the target search times \cite{slutsky+mirny_04, Coppey+al_04, Sokolov+al_05, Lomholt+al_05}. While our model does not examine all of the proposed mechanisms and makes approximations about the geometry of the TF and the DNA, we are ultimately only interested in what happens locally around the specific site as opposed to computing global properties such as target search time statistics. As a result, while rare but long range excursions might be important in search arguments, they should not significantly affect the noise as long as they are not the dominant form of all bulk transfers; if they are, one would have the case discussed in Section V.A of perfect mixing upon dissociation. Moreover,  the conclusions presented here reflect the basic differences in the diffusive processes in one and three dimensions, in particular the high probability of returning to the origin in 1D diffusion, and are therefore not a consequence of the detailed assumptions about the TF-DNA interaction, as is clear from the ``back of the envelope'' arguments presented in the Section II.

The diffusive contribution to the total noise in gene expression has only recently been recognized as significant. Metzler discusses the concept of ``interaction volume'' around the regulatory site on the DNA and studies the probability that transcription factors will enter this volume in the case of the $\lambda$-phage infecting \emph{Escherichia coli} bacterium, concluding that the spatial fluctuations can be important in genetic circuits \cite{Metzler_01}. Holcman et al study the master-diffusion equation in the context of signaling molecules binding to and unbinding from ion channels \cite{Holcman+Schuss_01}. The analysis of Bicoid/Hunchback system in the development  of the fruit fly \emph{Drosophila melanogaster} shows that the measured signatures in the noise in Hunchback expression are consistent with diffusive fluctuations in Hunchback's regulator, Bicoid \cite{gregor+al_07b,tkacik+al_07}. Van Zon and coworkers \cite{vanZon+al_06} study the diffusional component of the noise in Green's function reaction dynamics (GFRD) stochastic simulations and conclude by noting that 1D sliding along the DNA could have important effects on the noise power spectra. 
All these results point to the basic physics of diffusion as setting a limit to precision of a fundamental biological process, transcriptional regulation.  Our results indicate that this fundamental limit is not easily evaded by the (still largely unknown) complexities of protein motion along the DNA molecule.

\begin{acknowledgments}
We thank T. Gregor for helpful comments on the manuscript.
This work was supported in part by NIH grants P50 GM071508 and R01 GM077599,  by NSF Grant PHY--0650617,
and by the Burroughs Wellcome Fund Program in Biological Dynamics (GT). We thank the Center for Studies in Physics and Biology at Rockefeller University for its hospitality.
\end{acknowledgments}


\begin{thebibliography}{99}
\bibitem{intuition}
Several colleagues have presented this intuitive argument as a response to the discussion in Ref \cite{gregor+al_07b}, which claims that the observed precision of transcriptional regulation in the early fruit fly embryo is close the physical limit.  We don't know of anyone who has put this argument into print, however.  We take this opportunity to thank those who insisted on a more rigorous answer.
\bibitem{tavg}
In case of transcriptional control, where the regulated protein is the ``readout'' of the transcription factor concentration, the integration time is usually the regulated protein lifetime (minutes or more); during this time the specific site occupancy will have equilibrated, implying that many translocation-related steps (binding and unbinding transitions to/from the specific site and the sliding state) occur during one integration time.
\bibitem{boundlinear}
Diffusion-limited reaction rates generally take the form $k_+=4\pi D L$, where $L$ is the effective target size that is primarily determined by the largest \emph{linear} extent of the target. See Ref \cite{Berg+vonHippel_85} for details.
\end{thebibliography}
\end{document}